\documentclass[conference]{IEEEtran}
\IEEEoverridecommandlockouts
\usepackage{cite}
\usepackage{amsmath,amssymb,amsfonts}
\usepackage{algorithmic}
\usepackage{algorithm}
\usepackage{graphicx}
\usepackage{textcomp}
\usepackage{xcolor}
\usepackage{threeparttable}
\usepackage{pifont}
\usepackage{booktabs}
\usepackage{url}

\def\BibTeX{{\rm B\kern-.05em{\sc i\kern-.025em b}\kern-.08em
    T\kern-.1667em\lower.7ex\hbox{E}\kern-.125emX}}
\begin{document}

\title{End-to-End Transformer Acceleration Through Processing-in-Memory Architectures}

%\author{\IEEEauthorblockN{Anonymous Author(s)}
%}
\author{
Xiaoxuan Yang$^{*}$, Peilin Chen$^{*}$, Tergel Molom-Ochir$^{\dagger}$, and Yiran Chen$^{\dagger}$ \\
Email: \{xiaoxuan, peilin\}@virginia.edu, \{tm431, yiran.chen\}@duke.edu \\
$^{*}$Department of Electrical and Computer Engineering, University of Virginia, Charlottesville, VA, USA \\
$^{\dagger}$Department of Electrical and Computer Engineering, Duke University, Durham, NC, USA
}

\maketitle

\begin{abstract}
Transformers have become central to natural language processing and large language models, but their deployment at scale faces three major challenges. First, the attention mechanism requires massive matrix multiplications and frequent movement of intermediate results between memory and compute units, leading to high latency and energy costs. Second, in long-context inference, the key-value cache~(KV cache) can grow unpredictably and even surpass the model’s weight size, creating severe memory and bandwidth bottlenecks. Third, the quadratic complexity of attention with respect to sequence length amplifies both data movement and compute overhead, making large-scale inference inefficient. To address these issues, this work introduces processing-in-memory solutions that restructure attention and feed-forward computation to minimize off-chip data transfers, dynamically compress and prune the KV cache to manage memory growth, and reinterpret attention as an associative memory operation to reduce complexity and hardware footprint. Moreover, we evaluate our processing-in-memory design against state-of-the-art accelerators and general-purpose GPUs, demonstrating significant improvements in energy efficiency and latency. Together, these approaches address computation overhead, memory scalability, and attention complexity, further enabling efficient, end-to-end acceleration of Transformer models.
\end{abstract}


\section{Introduction}

Large language models~(LLMs) have achieved remarkable success in natural language processing tasks, such as question answering, code completion, and text summarization~\cite{naveed2025comprehensive, nijkamp2022codegen, yang2024harnessing, zhang2025systematic}. The transformer architecture, which is built upon attention mechanisms, serves as the foundation for LLMs~\cite{vaswani2017attention, radford2019language, devlin2019bert}. Specifically, the scaled dot-product attention~(SDA) relies on queries, keys, and values to compute contextual dependencies. To obtain stronger capabilities, LLMs have undergone rapid scaling in model size and context length~\cite{hooper2024kvquant, xiao2023smoothquant}.

Processing-in-memory~(PIM) architecture is considered a promising computing paradigm~\cite{chi2016prime, song2017pipelayer,yang2022research, chen2025optimizing} and has been widely used to accelerate convolutional neural networks~(CNNs), recurrent neural networks~(RNNs), generative adversarial network (GAN), graph processing, among others~\cite{long2018reram,song2018graphr,chen2018regan,chen2019zara,wang2021rerec,wang2020reboc,zheng2020mobilatice,yang2021multi,wu2024block,cheng2025autorac}. However, directly applying PIM architecture to accelerate Transformer-based LLMs faces many challenges due to the unique computational characteristics of SDA~\cite{yang2023improving,cong2024attentionlego,wolters2024memory}. 

\textit{Challenge 1: Intermediate Result Management}. Unlike the static weight matrices in CNNs, the query~(Q), key~(K), and value~(V) matrices in Transformer are dynamically generated at runtime. In resistive random-access memory-based~(ReRAM) PIM designs~\cite{chi2016prime}, these intermediate data must be frequently reprogrammed into the ReRAM arrays to enable subsequent Matrix-matrix Multiplication~(MatMul) such as $Q\times K^T$. However, programming a matrix into ReRAM is often performed column by column and introduces significant latency~\cite{sheu20114mb, yang2020retransformer}. In addition, the multi-head attention block repeatedly regenerates and rewrites these matrices, resulting in excessive write energy. This dynamic nature creates an inherent compute-write-compute~(CWC) dependency, as the MatMul cannot proceed until the intermediate data~($K^T$) are fully written into the ReRAM crossbars.

\textit{Challenge 2: KV Cache Memory Bottleneck}. Existing systems store the computed key and value within the SDA to skip redundant computations~\cite{jin2024compute, park2020optimus}. However, the KV cache has emerged as a dominant memory bottleneck in Transformer-based LLMs. Unlike static weights that are fixed after training, keys and values are dynamically generated during inference. In large-scale LLMs such as 540B PaLM, the KV cache can reach 3 TB~(3$\times$ of model's static weights) under a batch size of 512 and a context length of 2048~\cite{liu2024kivi}. Moreover, the KV cache size depends on the total context length and grows unpredictably. Existing accelerators~\cite{zeng2024flightllm, park2020optimus, qu2022dota} and GPUs store both static weights and KV cache in off-chip memory. Due to the auto-regressive decoder, the entire KV cache and weights must be repeatedly reloaded to generate each new token~\cite{chen2025titanus}. This will lead to massive data movement and energy consumption overhead.

\textit{Challenge 3: Quadratic attention \& long-context scaling.} Scaled dot-product attention’s similarity stage $QK^{\top}$ grows as $O(N^{2})$ in sequence length $N$. As contexts extend (e.g., $4\text{K}\!\rightarrow\!32\text{K}\!\rightarrow\!128\text{K}\!\rightarrow\!1\text{M}$), both arithmetic and more critically for PIM-based designs, data movement grow explosively: repeated key reads, large softmax working sets, and frequent on/off-chip transfers dominate latency/energy even with IO-aware tiling such as FlashAttention~\cite{dao2022flashattention}. In practice, the wall-clock profile shifts from compute-bound to traffic-bound as longer contexts overflow on-chip buffers, incur extra passes, and saturate bandwidth.

This work overcomes the above challenges with three major contributions. For \textbf{Challenge 1}, we propose a matrix decomposition technique to eliminate CWC dependency by avoiding frequent rewriting of intermediate results. Moreover, a new sub-matrix pipeline design for the SDA is introduced to further improve system throughput and power efficiency. For \textbf{Challenge 2}, we propose a software-hardware co-design that employs a cascade pruning-quantization~(CPQ) method to compress the KV cache on-the-fly, and transfers only the non-zero KV cache between the accelerator and off-chip memory to further reduce data movement. For \textbf{Challenge~3}, we reformulate attention as nearest-neighbor retrieval. A lightweight proxy guides computation toward promising matches, after which calibrated re-scoring yields sparse attention weights, cutting similarity operations and $V$ reads compared to dense $QK^{\top}$ while remaining compatible with our CPQ KV cache compression and sub-matrix pipeline.

\section{Background and motivation}
\label{section2}

\begin{figure}[tb]
    \centering
    \setlength{\abovecaptionskip}{0pt}
    \includegraphics[width=0.7\linewidth]{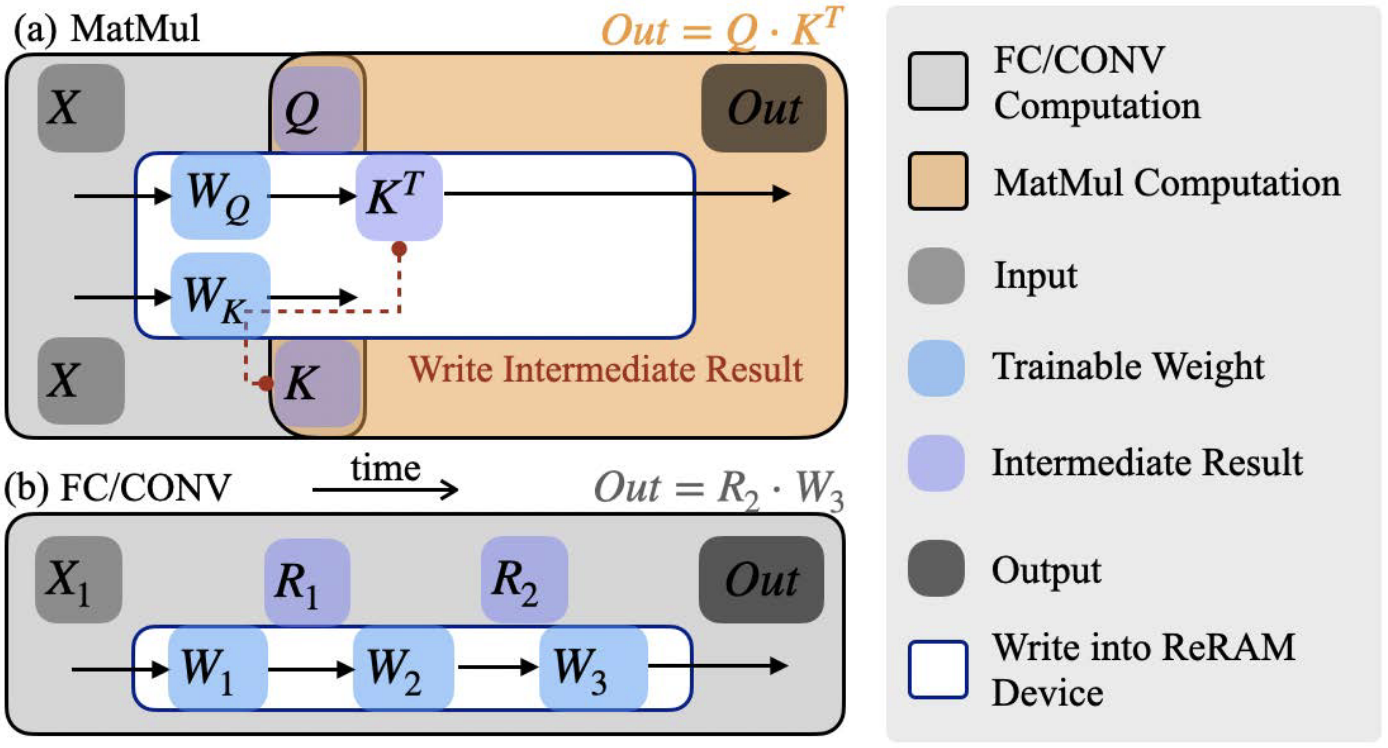}
    \caption{Comparison between FC/CONV layers and MatMul layer in ReRAM implementation~\cite{yang2020retransformer}: (a) MatMul layers, (b) FC/CONV layer.}
    \vspace{-18pt}
    \label{fig1}
\end{figure}

\subsection{Compute-Write-Compute Dependency for Attention Block}
\label{section2_1}

Unlike traditional CNNs, where the computation of fully-connected~(FC) and convolutional~(CONV) layers can be directly pipelined across inputs and weights, the attention block in Transformer introduces a compute-write-compute~(CWC) dependency due to its unique computational characteristics. As shown in Fig.~\ref{fig1}(a), the computation of $Q\times K^T$ requires intermediate results~($K^T$) that must be generated and then written into the ReRAM arrays before subsequent multiplication can continue. CWC dependency not only increases latency but also incurs significant write overhead on ReRAM-based PIM systems. In contrast, FC/CONV layers can avoid this issue because outputs can be directly propagated as inputs to the next layer without storing intermediate results~(Fig.~\ref{fig1}(b)). Therefore, the CWC dependency becomes a unique challenge for the ReRAM-based PIM architecture.

\subsection{Key-Value Cache Bottleneck for Transformer Models}

To mitigate the overhead of the KV cache in LLMs, prior algorithm work ThinK~\cite{xu2024think} performs key-only pruning and ignores the most recent and newly generated key cache. KIVI~\cite{liu2024kivi} and KVQuant~\cite{hooper2024kvquant} apply group-wise quantization to reduce KV cache bit-width~(e.g., 2-bit or 3-bit) with limited accuracy degradation. These methods achieve limited pruning ratios and do not consider the layer-wise variations in KV cache sensitivity to pruning and quantization. Moreover, prior hardware designs~\cite{ham20203, ham2021elsa, qu2022dota, zeng2024flightllm} explore approximate attention mechanisms, low-rank linear transformation, and structure sparsity~(e.g., N$:$M pattern) to reduce the quadratic complexity of the self-attention mechanism. A key limitation of previous work is the substantial energy overhead incurred by repeatedly transferring KV cache between off-chip memory and the accelerator. Therefore, we are motivated to design dedicated hardware to compress the KV cache on-the-fly, since KV cache cannot be pre-optimized like static weights.

\subsection{Attention Complexity and Prior Mitigations}
Self-attention forms $QK^{\top}$ and normalizes the result; with sequence length $N$, the similarity stage scales as $O(N^{2})$ and becomes \emph{IO-bound} even with IO-aware exact kernels like FlashAttention~\cite{dao2022flashattention}. Operator-level approximations (low-rank, structured sparsity, approximate attention) reduce MACs but introduce indexing/metadata movement and still leave substantial KV traffic~\cite{ham20203,ham2021elsa,qu2022dota}. Orthogonally, KV-cache compression methods (quantization/pruning) shrink footprint and transfers~\cite{hooper2024kvquant,liu2024kivi,xu2024think}, yet remain layer/stage sensitive and grow with context. On PIM-based designs, these residual effects are further amplified: (i) the \emph{compute–write–compute} dependence forces costly intermediate writes~(Sec.~\ref{section2_1}); (ii) long contexts overflow on-chip buffers; and (iii) repeated KV movement dominates energy~(Sec.~II-B). We therefore combine (a) an algebraic decomposition with a sub-matrix pipeline~(Sec.~III), (b) an on-the-fly CPQ compression strategy for KV cache with dedicated hardware support~(Sec.~IV), and (c) an \emph{associative retrieval} view that focuses similarity computation instead of dense $QK^{\top}$~(Sec.~V).

\section{Matrix Decomposition and Sub-Matrix Pipeline in SDA}
\begin{figure}[tb]
    \centering
    \setlength{\abovecaptionskip}{0pt}
    \includegraphics[width=0.85\linewidth]{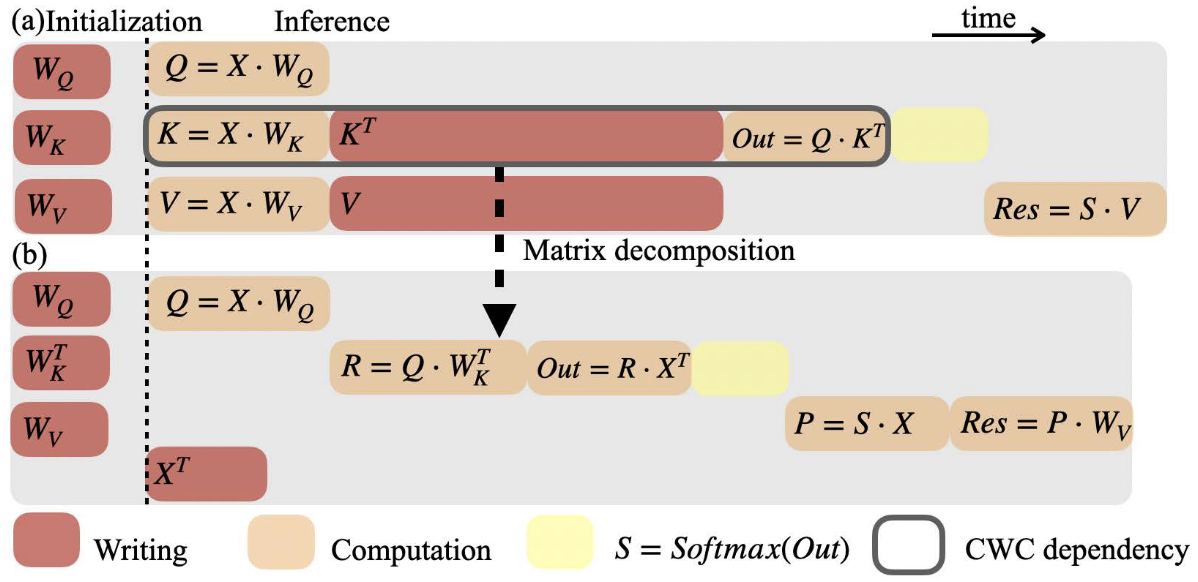}
    \caption{Illustration of eliminating the CWC dependency in scaled dot-product attention~(SDA)~\cite{yang2020retransformer}.}
    \vspace{-15pt}
    \label{fig2}
\end{figure}
\subsection{Matrix Decomposition}
As discussed in Section~\ref{section2_1}, the CWC dependency between the linear projection layer and the subsequent MatMul is a critical challenge. In the standard formulation~(Fig.~\ref{fig2}(a)), query $Q{=}X\times W_Q$, key $K{=}X\times W_K$, and value $V{=}X\times W_V$ must first be generated. To perform the MatMul of $Out{=}Q\times K^T$, $K^T$ then needs to be stored in the ReRAM crossbar before the computation can proceed. To address this problem, we introduce an algebraic reformulation of the MatMul in SDA:
\begin{equation*}
    Out=Q\times K^T=Q\times (X\times W_K)^T=(Q\times W^T_K)\times X^T.
\end{equation*}
Instead of directly calculating $Out{=}Q\times (X\times W_K)^T$, we decompose the MatMul between $Q$ and $K$ into two cascaded multiplications~($Out{=}(Q\times W^T_K)\times X^T$), as shown in Fig.~\ref{fig2}(b). This decomposition removes the need to write $K^T$
into the ReRAM crossbar. The system can pre-load $W_Q$ and $W^T_K$ into two crossbars, compute $Q{=}X\times W_Q$, then obtain $R{=}Q\times W^T_K$, and finally complete $Out{=}R\times X^T$. By shifting the dependency from storing $K^T$ to directly reusing $X^T$, the long programming latency of writing $K^T$ into the crossbar is avoided. Moreover, this reformulation can be extended to subsequent steps. For example, the MatMul between $S$ and $V$ in SDA can also be optimized similarly. One practical consideration is that both $W_K$ and $W_V$ computation threads may require access to $X$
or $X^T$. To minimize the overhead of maintaining duplicate data, $Out{=}R\times X^T$ and $P{=}S\times X$ are executed sequentially using a single copy of $X$ in a dual-access design of ReRAM crossbars~\cite{li2018rc}. 

\subsection{Sub-Matrix Pipeline Design}
\begin{figure}[tb]
    \centering
    \setlength{\abovecaptionskip}{0pt}
    \includegraphics[width=0.7\linewidth]{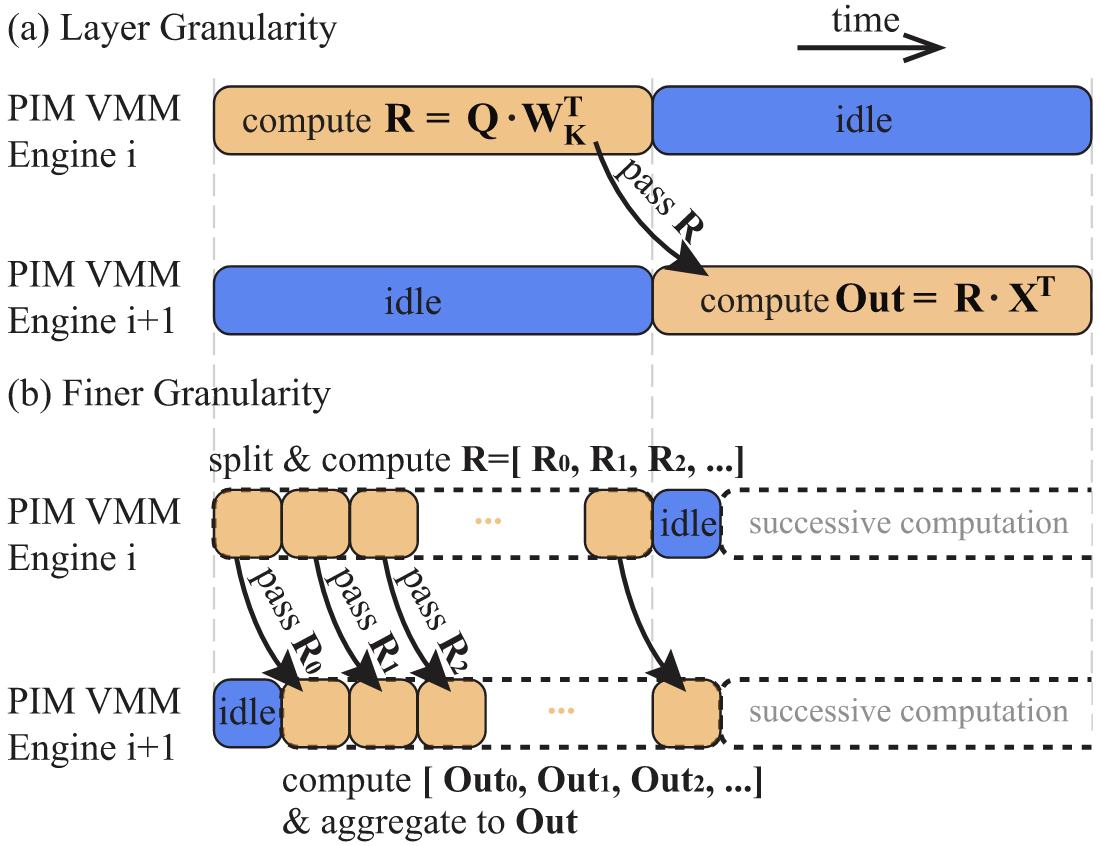}
    \vspace{-4pt}
    \caption{(a) Layer-level pipeline and (b) Sub-matrix level pipeline~\cite{yang2020retransformer}.}
    \vspace{-12pt}
    \label{fig3}
\end{figure}
\begin{figure}[tb]
    \centering
    \setlength{\abovecaptionskip}{0pt}
    \includegraphics[width=\linewidth]{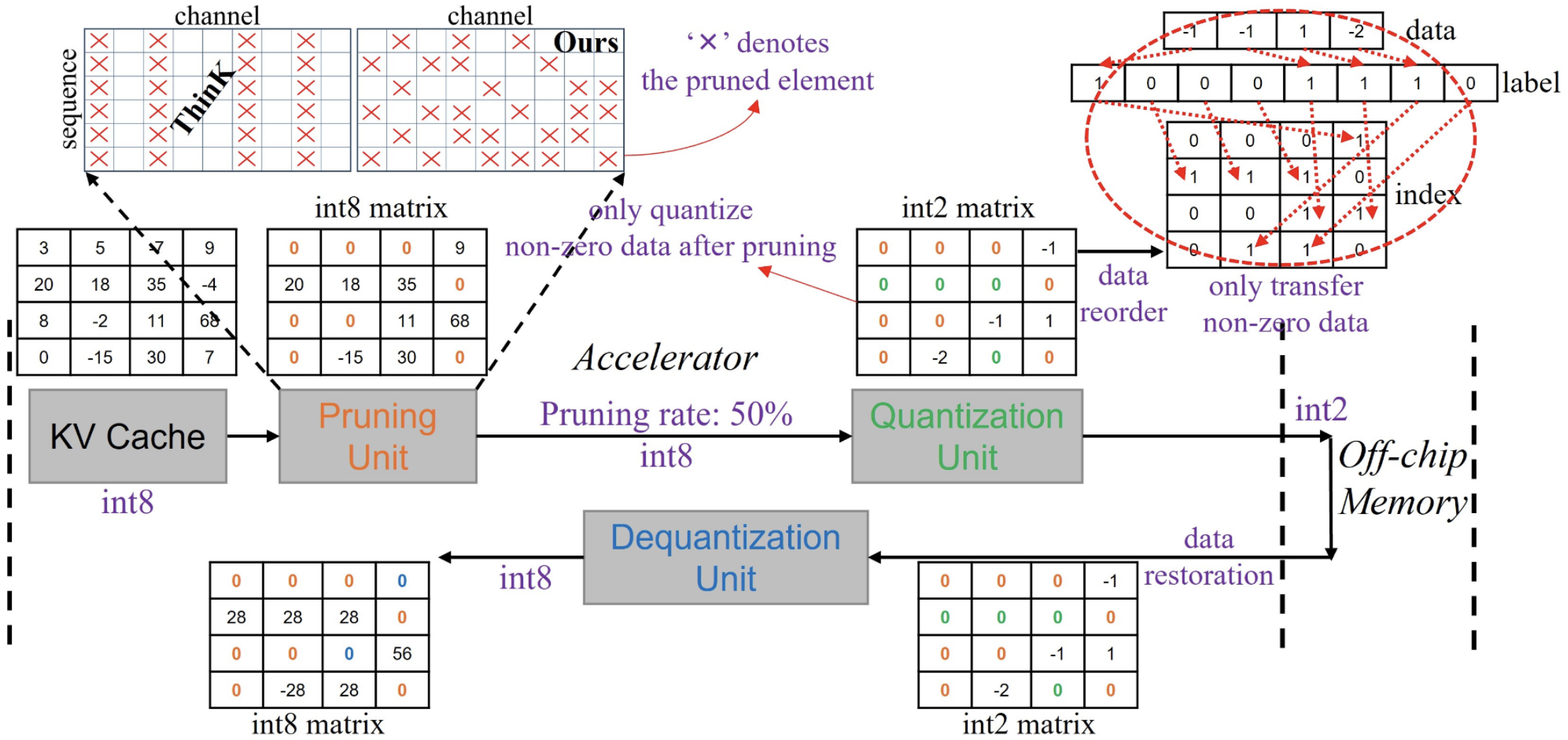}
    \caption{Overview of the CPQ compression strategy~\cite{chen2025titanus}.}
    \vspace{-18pt}
    \label{fig4}
\end{figure}
Conventional PIM-based accelerators~\cite{song2017pipelayer} exploit layer-level pipeline execution to achieve high inference or training throughput. However, layer granularity can lead to under-utilization of ReRAM crossbars. As shown in Fig.~\ref{fig3}(a), when two sequential MatMuls $R{=}Q\times W^T_K$ and $Out{=}R\times X^T$ are executed at the layer-level pipeline manner, one crossbar remains idle while the other is active, resulting in low hardware utilization. To address this issue, we introduce a new sub-matrix pipeline design that operates at a finer granularity. Specifically, the input matrices are divided into smaller segments that can be processed successively. For example, as shown in Fig.~\ref{fig3}(b), $Q$ is split into sub-vectors, each multiplied with $W^T_K$ to produce partial rows of $R$. These partial results are then streamed into the subsequent multiplication with $X^T$. In this way, the two sequential MatMuls will overlap in time, thus keeping both crossbars active while minimizing idle time.

\section{Cascade KV Cache Pruning-Quantization}

\begin{figure}[tb]
    \centering
    \setlength{\abovecaptionskip}{0pt}
    \includegraphics[width=0.88\linewidth]{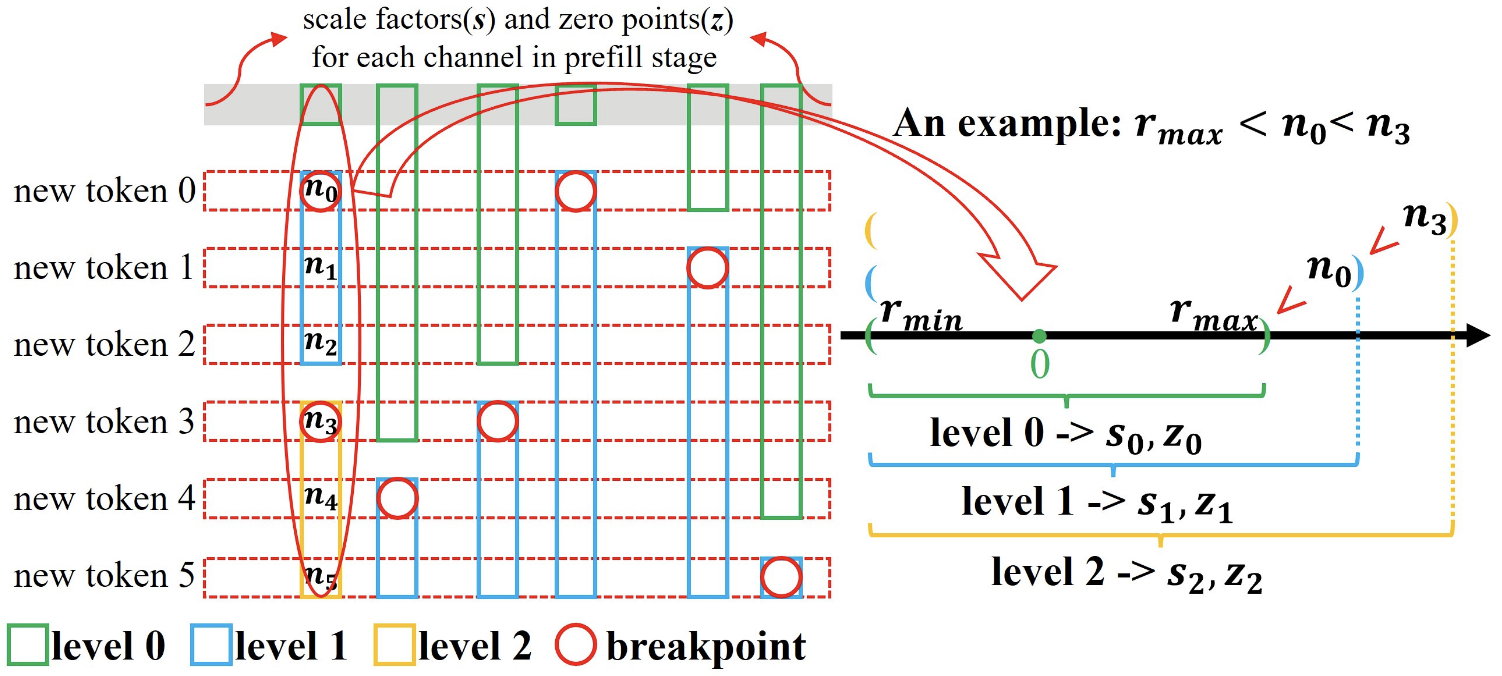}
    \caption{Hierarchical quantization extension strategy~\cite{chen2025titanus}.}
    \vspace{-20pt}
    \label{fig5}
\end{figure}

\textbf{Software-level Optimization.} An important direction for reducing the KV cache in LLMs is to utilize pruning and quantization compression techniques~\cite{liu2024kivi, hooper2024kvquant, xu2024think}. Instead of treating them separately, we propose a cascade pruning-quantization strategy~(Fig.~\ref{fig4}) that first prunes redundant elements and then applies quantization only to the remaining non-zero data. More importantly, we will reorder the quantized KV cache and transfer only the non-zero KV cache between off-chip memory and the accelerator to reduce data movement. \textit{For the pruning part,} unlike the ThinK~\cite{xu2024think}, we leverage a finer granularity strategy to prune the unimportant elements of KV cache~(upper-left side of Fig.~\ref{fig4}). Moreover, pruning is applied not only to the prefill stage but also to the decode stage, enabling continuous savings across the inference process. \textit{For the quantization part,} we adopt the per-channel quantization~(PCQ) method since the overhead of token-wise quantization is significantly larger than PCQ~\cite{chen2025titanus}. However, there is a non-independent PCQ~(NiPCQ) issue that re-quantizes all channels when a newly generated token arrives. We introduce a hierarchical quantization extension strategy~(HQE) to solve this challenge~(Fig.~\ref{fig5}). Specifically, instead of relying on a single set of quantization parameters for each channel, HQE progressively extends the tolerance range~(TR) during decoding by introducing new quantization levels when tokens fall outside the initial range. As shown in Fig.~\ref{fig5}, scale factors and zero points are first determined in the prefill stage. The newly generated tokens are then monitored, and the parameters of the new level will be created only when a token exceeds the TR of the previous level. This method ensures that each token is quantized once, thereby reducing computation overhead.

\begin{figure*}[tb]
    \centering
    \setlength{\abovecaptionskip}{0pt}
    \includegraphics[width=0.69\linewidth]{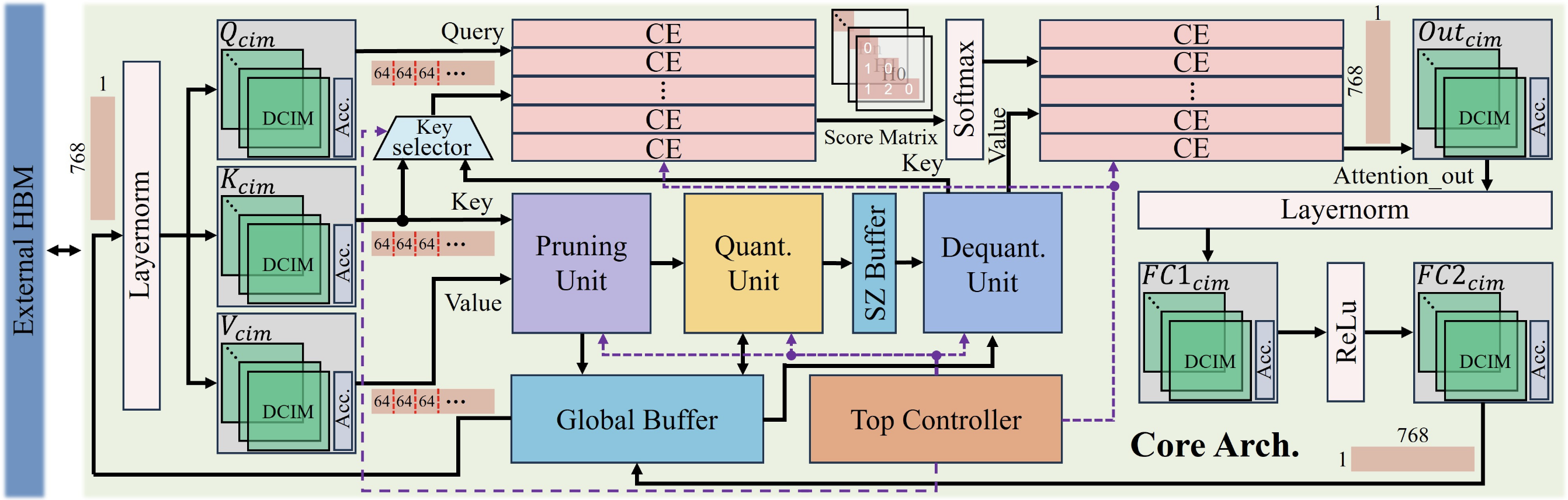}
    \caption{Core-level overall architecture~\cite{chen2025titanus}. CE and SZ denote the computing engine and scale-zero buffer, respectively.}
    \vspace{-18pt}
    \label{fig6}
\end{figure*}

\textbf{Hardware-level Optimization.} We design dedicated hardware to support the proposed CPQ compression strategy on-the-fly, as the KV cache is dynamically generated during inference. Fig.~\ref{fig6} presents the overall architecture of the core-level design. Note that multiple cores are utilized to perform the end-to-end LLM inference. The digital CIM~(DCIM)~\cite{chih202116} blocks~($Q_{cim}$, $K_{cim}$, $V_{cim}$, $Out_{cim}$, $FC1_{cim}$, and $FC2_{cim}$) are used to store trainable matrices such as query, key, and value projections. We also incorporate the computing engine~(CE) that leverages sparsity introduced by the CPQ method. The CE operates in parallel with the number of attention heads and dynamically activates processing units according to the input data size. In addition, several specialized units are employed to support the CPQ method. Pruning unit~(PU) removes the insignificant elements of the KV cache on-the-fly and forwards only non-zero data to the following modules. Quantization unit~(QU) is designed to support the proposed HQE method. Specifically, QU applies low-bit quantization to the pruned non-zero KV cache and enables the channel-wise monitoring mechanism at runtime. Finally, the dequantization unit~(DQU) reconstructs the compressed KV cache using the auxiliary data~(e.g., label and index in Fig.~\ref{fig4}). The experimental result shows that our software-hardware co-design achieves $159.9\times$~($49.6\times$) and $34.8\times$~($29.2\times$) energy efficiency (throughput) compared to Nvidia A100 GPU and FlightLLM~\cite{zeng2024flightllm} under end-to-end OPT-6.7B~\cite{zhang2022opt} inference.

\section{Attention as nearest-neighbor retrieval}
The attention mechanism in Transformers can be fundamentally understood as a nearest-neighbor retrieval operation. Given query $\mathbf{q}_i$ and key vectors $\{\mathbf{k}_j\}_{j=1}^N$, attention weights $A_{ij} = \exp(\mathbf{q}_i^T \mathbf{k}_j / \sqrt{d_k}) / \sum_{l=1}^N \exp(\mathbf{q}_i^T \mathbf{k}_l / \sqrt{d_k})$ represent similarity scores that effectively perform soft selection of the most relevant keys. This can be reformulated as a two-stage process: (1) similarity computation to find candidate matches, and (2) attention weighting to aggregate corresponding values. Recent work has shown that attention patterns exhibit sparsity, with most weights being negligibly small \cite{khandelwal2019generalization, guu2020retrieval}, making nearest-neighbor retrieval a natural paradigm for attention computation. Furthermore, focusing on the top-$K$  keys reduces computational complexity from $O(N^2)$ to $O(K \cdot N)$ \cite{he2021efficient}.

Content-Addressable Memory (CAM) provides an efficient hardware primitive for nearest-neighbor search operations. CAMs can search for content matching a given query pattern, aligning perfectly with attention's need to find similar keys. FeFET-based multi-bit CAMs have demonstrated in-memory nearest-neighbor search with L2 distance metrics~\cite{kazemi2020fefet_mcam}, while flash-based CAMs show efficient content retrieval capabilities~\cite{yang2023flashcam}. CAM integration with PIM architectures offers reduced data movement, parallel processing across multiple CAM arrays, and significant energy efficiency gains compared to digital matrix multiplication~\cite{laguna2022imtransformer}. The CAM-based approach synergizes effectively with our previous techniques: matrix decomposition can store decomposed matrices directly in CAM arrays, CPQ KV cache compression naturally reduces keys stored in CAMs while quantization reduces precision requirements, and sub-matrix pipelines can be enhanced with CAM-based nearest-neighbor searches. Future work could explore adaptive CAM configurations for structured sparsity patterns~\cite{johnson2019billion}, specialized CAM designs for different attention head types~\cite{izacard2022few}, hierarchical architectures for long-context scaling~\cite{borgeaud2022improving}, and integration with emerging memory~\cite{manea2024gaincell}.

\section{Conclusion}
\label{section5}

This paper proposes the PIM-based designs that jointly optimize computation, KV cache bottleneck, and data movement for LLMs. Through matrix decomposition and sub-matrix pipeline, on-the-fly KV cache compression, and associative retrieval attention reformulation, we eliminate the dependency in attention block, manage data movement growth, and reduce the quadratic complexity of SDA. These techniques together offer a scalable path toward efficient deployment of LLMs.

\section*{Acknowledgement}
T. Molom-Ochir and Y. Chen acknowledge the support from NSF 2328805, 2112562, and DOE DE-SC0026254.

\bibliographystyle{ieeetr}
\bibliography{citations}

\end{document}